\documentclass{aip-cp}

\usepackage[numbers]{natbib}
\usepackage{units}
\usepackage{rotating}
\usepackage{graphicx}
\usepackage{color}

\newcommand{\eV}{e\kern-0.075emV}
\newcommand{\keV}{ke\kern-0.125emV}
\newcommand{\MeV}{Me\kern-0.1emV}
\newcommand{\GeV}{Ge\kern-0.1emV}

\begin{document}

\title{Baryon Spectroscopy - Recent Results from the CBELSA/TAPS Experiment}

\author[aff1]{Jan Hartmann\corref{cor1}}
\author{for the CBELSA/TAPS collaboration}

\affil[aff1]{Helmholtz-Institut f\"ur Strahlen- und Kernphysik, Universit\"at Bonn, Germany}
\corresp[cor1]{Corresponding author: hartmann@hiskp.uni-bonn.de}

\maketitle

\begin{abstract}
One of the remaining challenges within the standard model is to gain a good understanding of QCD in the non-perturbative regime. One key step toward this aim is baryon spectroscopy, investigating the spectrum and the properties of baryon resonances. To get access to resonances with small $\pi N$ partial width, photoproduction experiments provide essential information. In order to extract the contributing resonances, partial wave analyses need to be performed. Here, a complete experiment is required to unambiguously determine the contributing amplitudes. This involves the measurement of carefully chosen single and double polarization observables. 
The CBELSA/TAPS experiment with a longitudinally or transversely polarized target and an energy tagged, linearly or circularly polarized photon beam allows the measurement of a large set of polarization observables. Due to its good energy resolution, high detection efficiency for photons, and the nearly complete solid angle coverage it is ideally suited for the measurement of photoproduction of neutral mesons decaying into photons. 
Recent results for various double polarization observables in $\pi^0$ and $\eta$ photoproduction and their impact on the partial wave analysis are discussed.
\end{abstract}

\section{INTRODUCTION}
The spectrum of excited nucleon states reflects the dynamics of QCD in the non-perturbative regime. It has been studied for many years using $\pi$ beams. However, the spectrum of known nucleon resonances is in conflict with predictions from quark models \cite{isgur77,loring01}. Most obvious is the missing resonance problem, the fact that more states are predicted by the models at higher masses than have been observed experimentally. But also the ordering of excited states with positive and negative parity is partly in disagreement, the most prominent example being the $N(1440)\,1/2^+$ which is predicted by most quark models to be heavier than the $N(1535)\,1/2^-$. QCD calculations on the lattice \cite{edwards11}, though using unphysically large quark masses, yield a similar pattern as the non-relativistic quark model. Measuring the properties of the known resonances more precisely and searching for the missing resonances is essential to understand the discrepancies between theory and experiment.

Photoproduction experiments allow access to resonances with small $\pi N$ couplings and therefore have great potential to observe the missing resonances. The contributing resonances are extracted from the measured data in a partial wave analysis (PWA). In order to do this in an unambiguous way, a complete experiment \cite{chiang97} is needed, which requires the measurement of polarization observables. In this paper, the measurement of single and double polarization observables accessible with linearly polarized beam and a transversely polarized target are reported. They complement our published results with a longitudinally polarized target and linearly \cite{thiel12} and circularly \cite{gottschall14} polarized beam.

\section{EXPERIMENTAL SETUP}
The data presented were obtained with the CBELSA/TAPS experiment at ELSA \cite{hillert06}. The linearly polarized photon beam was produced from the incident $\unit[3.2]{\GeV}$ electron beam via coherent bremsstrahlung off a carefully aligned diamond crystal \cite{elsner09}. The coherent edge was set to $E_{\gamma} = \unit[950]{\MeV}$, resulting in a maximumum polarization of $65\%$ at $E_{\gamma} = \unit[850]{\MeV}$. The electrons passed through a magnet hitting a tagging hodoscope which defined the energy of the bremsstrahlung photons. The photon beam impinged on a frozen spin butanol target \cite{bradtke99} providing transversely polarized protons with an average target polarization of $74\%$.

\begin{figure}[t]
	\centering
	\includegraphics[width=.75\textwidth]{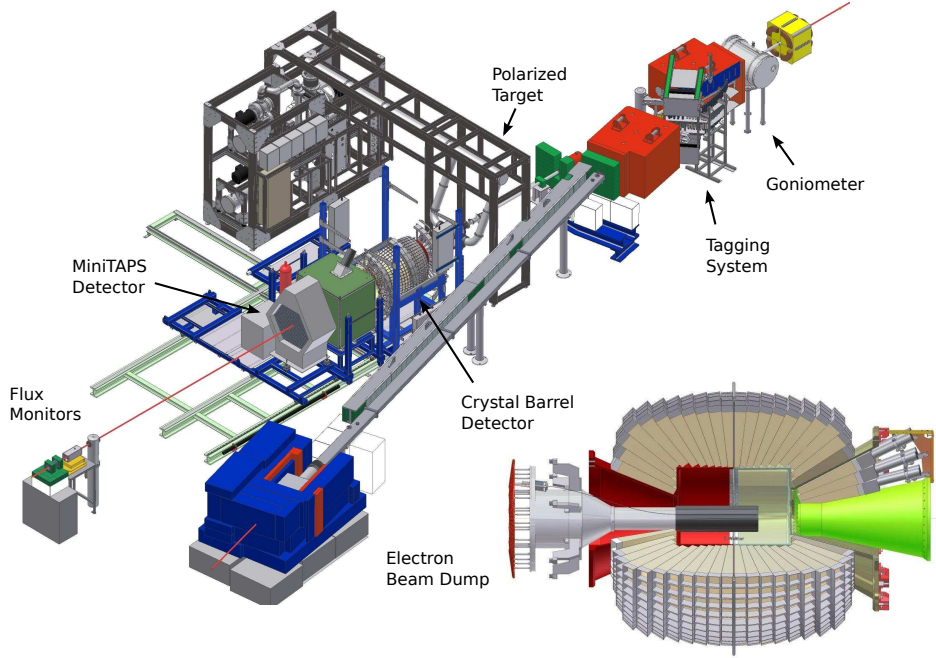}
	\label{fig:setup}
	\caption{The experimental setup of the CBELSA/TAPS experiment.}
\end{figure}

The detector system, which is shown in Figure~\ref{fig:setup}, consisted of two electromagnetic calorimeters, the Crystal Barrel \cite{aker92} and the MiniTAPS detector \cite{novotny91}, together covering the polar angle range from $1^\circ$ to $156^\circ$ and the full azimuthal angle. For charged particle identification, a three-layer scintillating fiber detector \cite{suft05} surrounding the target, and plastic scintillators in forward direction could be used. The detector setup provides a high detection efficiency for neutral particles and is therefore ideally suited to measure single and double polarization observables in reactions with neutral final states.

\section{DATA ANALYSIS}
\begin{figure}[ht]
	\centering
	\hspace{-0.03\textwidth}
	\includegraphics[width=.25\textwidth]{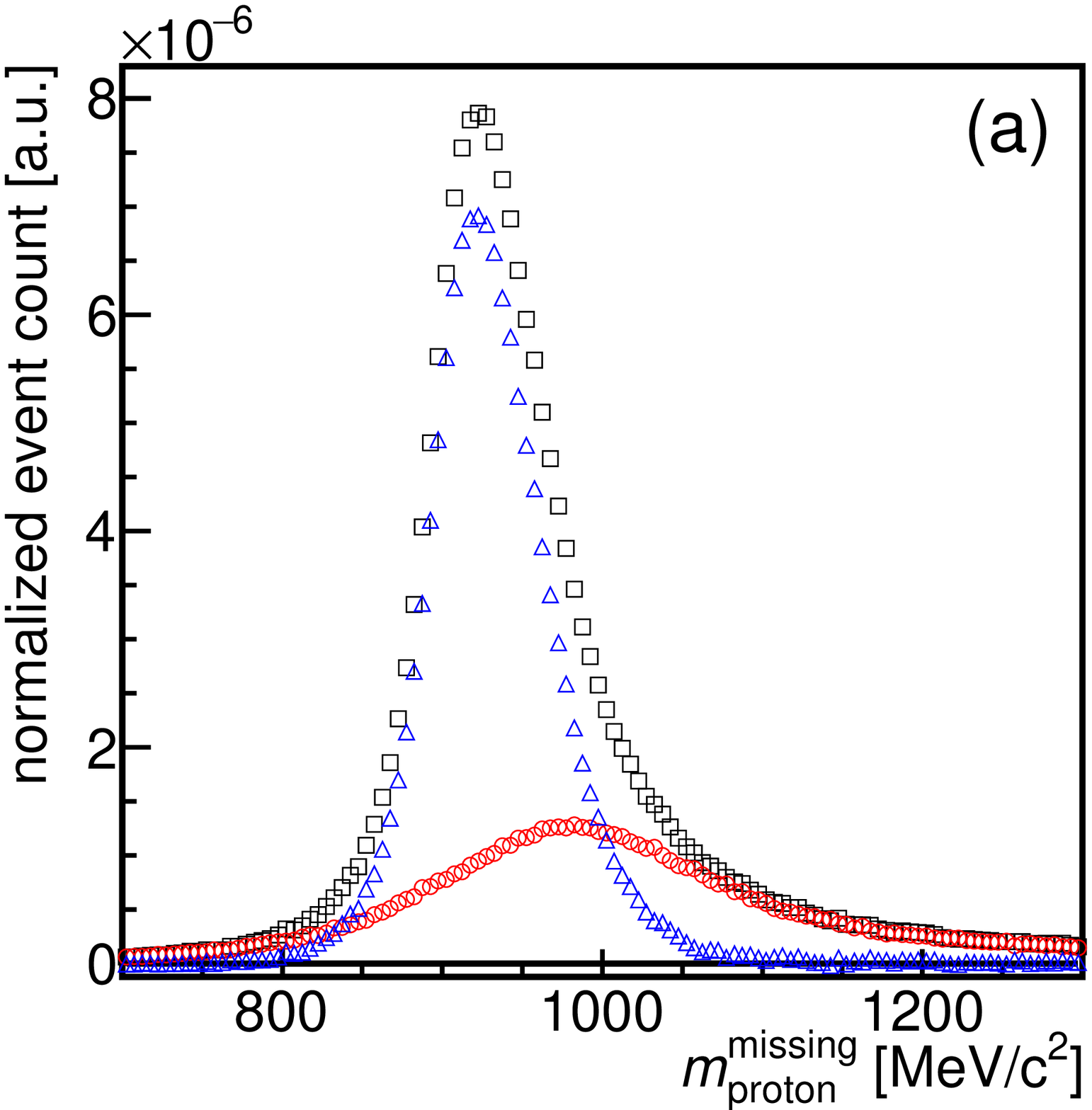}
	\hspace{-0.01\textwidth}
	\includegraphics[width=.25\textwidth]{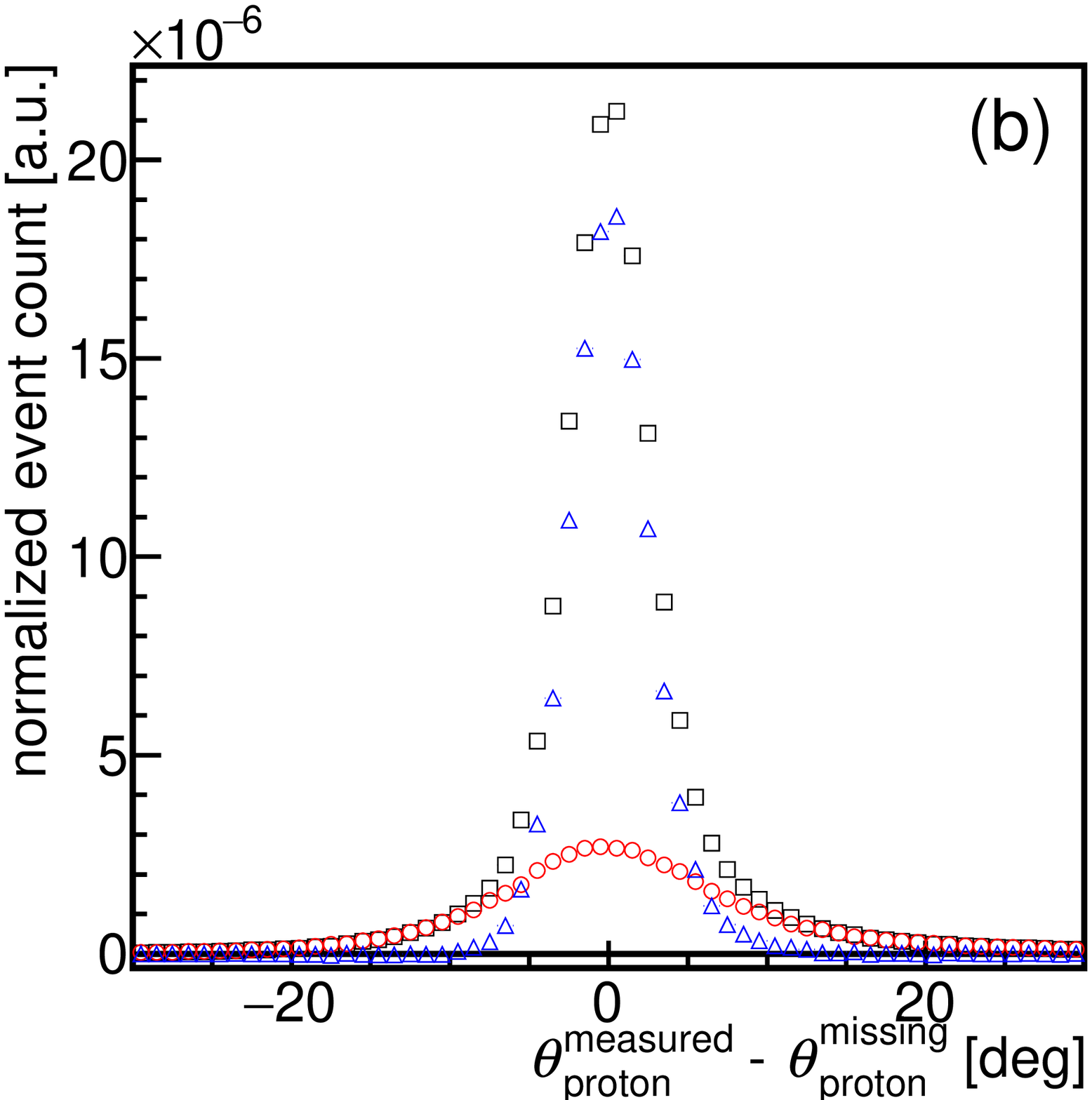}
	\hspace{-0.01\textwidth}
	\includegraphics[width=.25\textwidth]{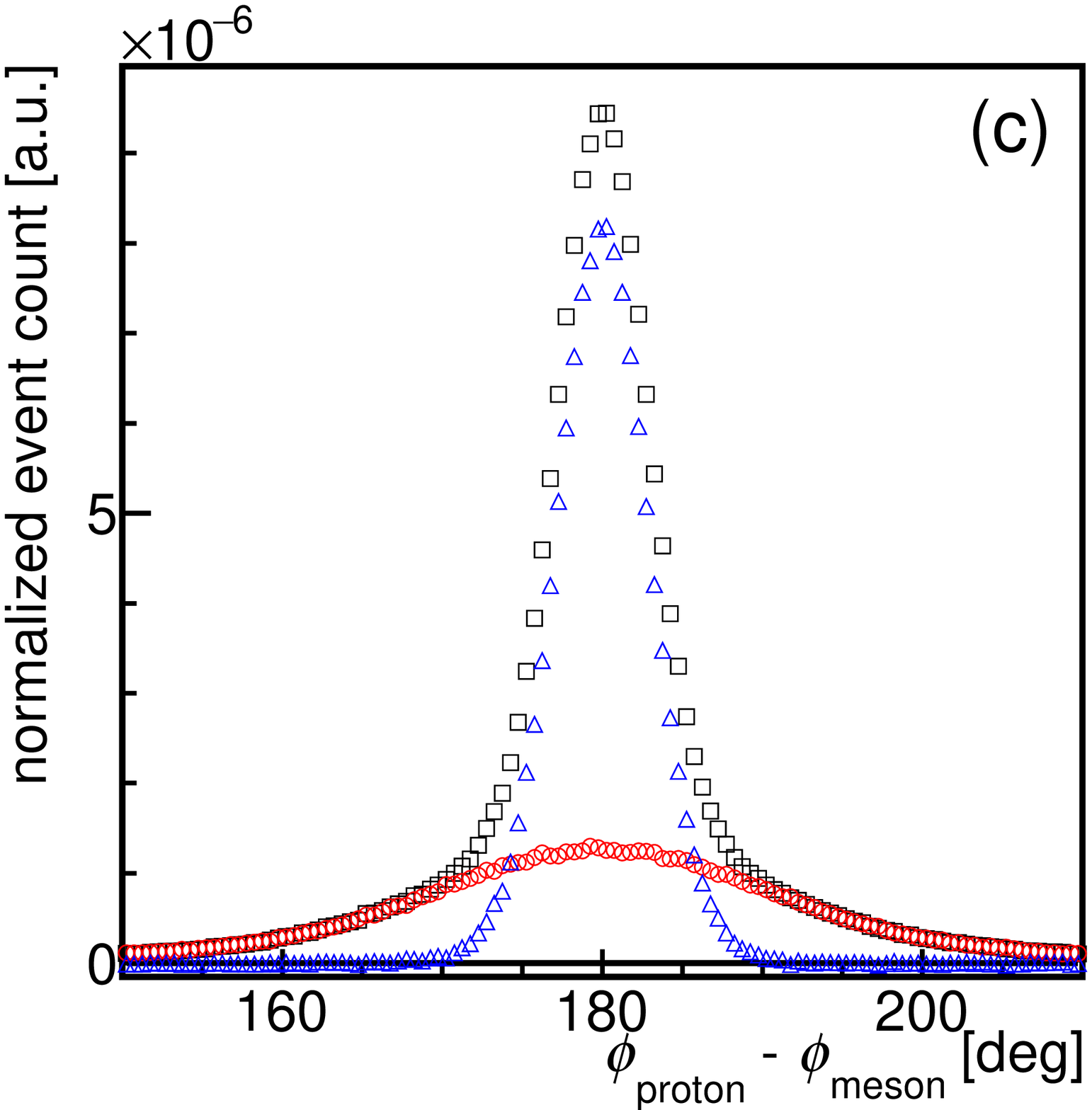}
	\hspace{-0.01\textwidth}
	\includegraphics[width=.25\textwidth]{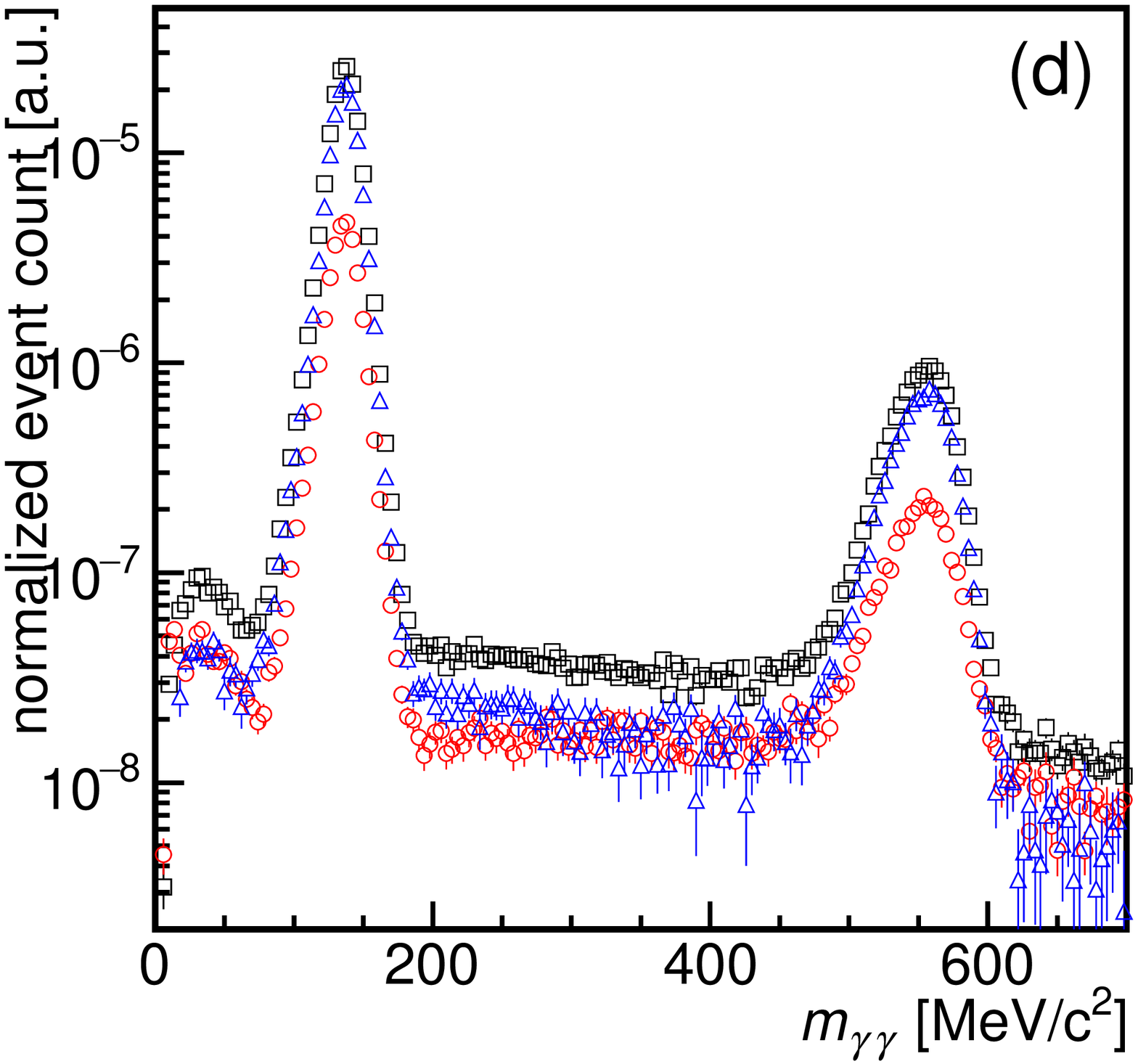}
	\label{fig:cuts}
	\caption{(a) The missing mass distribution, with the proton as the missing particle, (b) the polar angle difference of measured and missing proton, (c) the azimuthal angle difference of meson and proton, and (d) the $\gamma\gamma$ invariant mass distribution. The distributions are shown---after all other cuts discussed in the text are applied---for butanol ({\scriptsize$\square$}), carbon({\color{red}$\circ$}), and their difference ({\color{blue}\scriptsize$\triangle$}).}
\end{figure}

To select events from reaction $\gamma p \to \gamma \gamma p$, only events with three distinct calorimeter hits were retained. All three possible combinations were treated as $\gamma \gamma p$ candidates, with the proton being treated as a missing particle. A time coincidence was required between the tagger hit and the reaction products, and random time background was subtracted. Kinematic cuts were applied to ensure longitudinal and transverse momentum conservation within $\pm2\sigma$, and the missing mass had to agree with the proton mass within $\pm2\sigma$. Finally, events from reaction $\gamma p \to \pi^0 p$ or $\eta p$ were selected by requiring the $\gamma\gamma$ invariant mass to be within $\pm2\sigma$ of the $\pi^0$ or $\eta$ mass, respectively. Examples for the missing mass, angular difference, and $\gamma\gamma$ invariant mass distributions are shown in Figure~\ref{fig:cuts}. The final data sample contains a total of 1.7 million $\pi^0 p$ and 170 thousand $\eta p$ events. The background contamination below the $\pi^0$ peak in the $\gamma\gamma$ invariant mass spectrum is less than $1\%$ for all energies and angles, for the $\eta$ it is below $2\%$.

Since a butanol target was used, not only reactions off polarized and unpolarized free protons contribute to the selected event sample, but also reactions occurring off the bound unpolarized nucleons of the carbon and oxygen nuclei. The measured target polarization $p_t$ is therefore diluted by a factor $d$. Additional measurements using a carbon foam target were performed to determine the effective dilution factor $d$ as a function of the beam energy $E_{\gamma}$ and the angle $\theta$ of the produced meson in the center-of-mass frame:
\begin{equation}
	d(E_\gamma, \cos\theta) = \frac{N_\mathrm{free}}{N_\mathrm{butanol}} = \frac{N_\mathrm{butanol} - N_\mathrm{carbon}}{N_\mathrm{butanol}},
\end{equation}
which assumes that the nucleons bound in carbon and oxygen show the same response to the impinging photons. The carbon foam target had the same size as the butanol target and approximately the same area density as the carbon and oxygen part in the butanol. The carbon target replaced the butanol target in the frozen spin cryostat to match the experimental conditions of the butanol measurement as closely as possible. The carbon data was normalized to the butanol data in a kinematic region where no contribution from free protons is expected. The missing mass and angular difference distributions in Figure~\ref{fig:cuts} are smeared out for the carbon data because of the unknown Fermi momentum in the initial state. The difference between the butanol and the carbon data yields the free proton results. For further details on the dilution factor determination see Ref.~\cite{hartmann15}.

With a linearly polarized photon beam and a transversely polarized target the distribution of events $N$ as a function of the azimuthal angle $\phi$ between the scattering plane and the photon polarization plane is given by
\begin{equation}
	\frac{N(\phi)}{N_0} = 1 - p_\gamma\,\Sigma_\mathrm{eff} \cos(2\phi) + d\,p_t\,T \sin(\phi-\alpha) - d\,p_t\,p_\gamma\ \bigl[ P\cos(2\phi)\sin(\phi-\alpha) - H\sin(2\phi)\cos(\phi-\alpha) \bigr],
\end{equation}
where $\alpha$ is the azimuthal angle between the target polarization vector and the photon polarization plane, $p_\gamma$ is the degree of linear beam polarization, and $p_t$ is the target polarization degree. The occuring polarization observables $\Sigma_\mathrm{eff}$ (which mixes the beam asymmetry from free and bound nucleons), $T$, $P$, and $H$ are determined, for each $(E_\gamma,\cos\theta)$ bin, from an event-based maximum likelihood fit \cite{hartmann16,diss:hartmann} to the measured azimuthal distribution of events. At energies above $E_\gamma = \unit[933]{\MeV}$, where $p_\gamma$ is small, only $T$ is determined.

Systematic uncertainties include the uncertainty in the degree of photon ($4\%$) and proton ($2\%$) polarization, in the dilution factor ($1\%$--$4\%$, due to the relative normalization of the carbon data), and an additional absolute uncertainty due to the remaining background contribution. Further details on the estimation of the systematic uncertainties can be found in Refs.~\cite{elsner09,bradtke99,diss:hartmann}.

\section{RESULTS}
\subsection{Reaction $\gamma p \to \pi^0 p$}
\begin{figure}[p]
	\centering
	\begin{minipage}{\textwidth}
		\includegraphics[width=\textwidth]{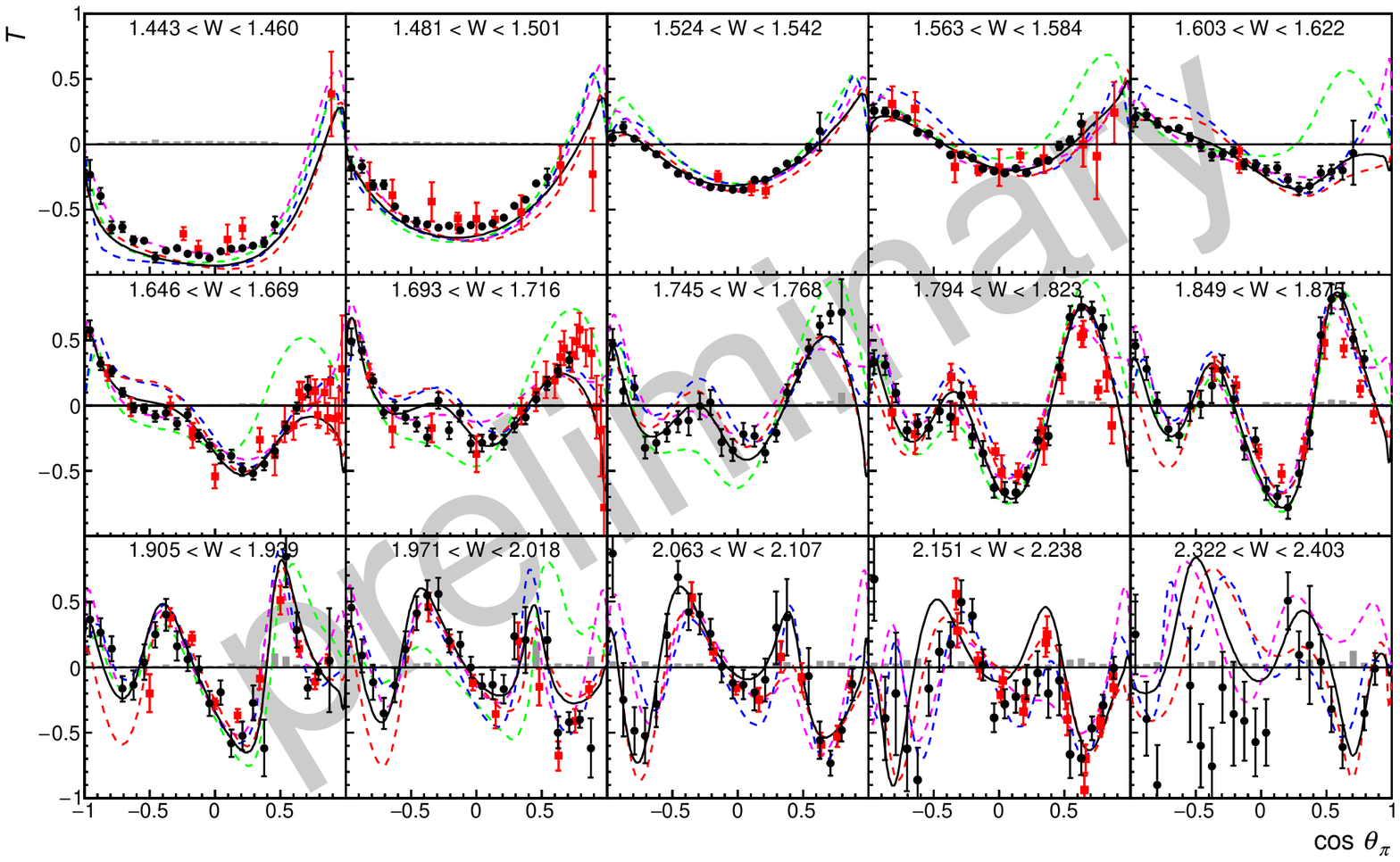}
		\includegraphics[width=\textwidth]{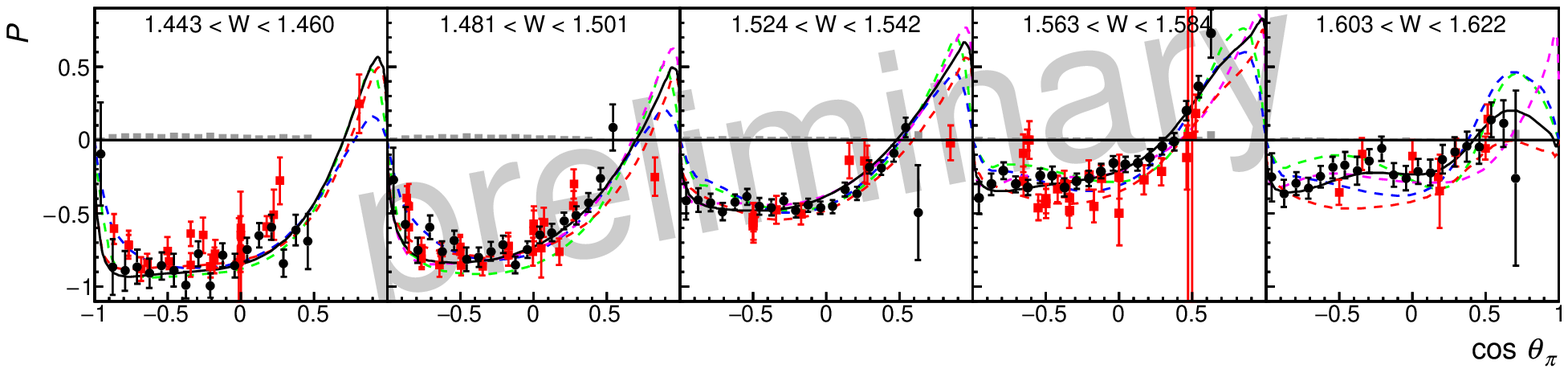}
		\includegraphics[width=\textwidth]{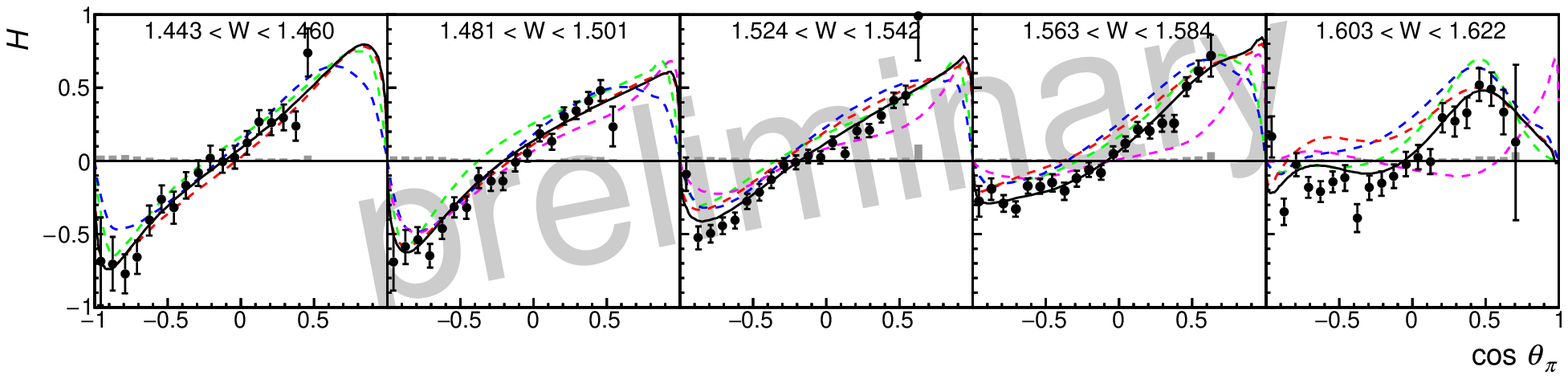}
	\end{minipage}
	\label{fig:results_pi0}
	\caption{The polarization observables $T$, $P$, and $H$ in the reaction $\gamma p \to \pi^0 p$ as a function of the scattering angle $\cos\theta_\pi$ and the $\gamma p$ invariant mass $W$ (in \GeV, only every second bin is shown here). The systematic uncertainty is shown as gray bars. References to earlier data (red points) are given in \cite{anisovich12}, refs. [49-71] therein. The solid black line represents the BnGa2014 fit \cite{hartmann15}. The data are compared to predictions (dashed curves) from BnGa2011-02 \cite{anisovich12} (red), MAID \cite{maid07} (green), SAID \cite{said12} (blue), and J\"uBo 2015 \cite{roenchen15} (magenta).}
\end{figure}

Results for the polarization observables $T$, $P$, and $H$ are shown in Figure~\ref{fig:results_pi0}. The data agree well with previously reported measurements but exceed the old data in precision and coverage in angles and energy. The agreement with predictions from BnGa2011 \cite{anisovich12}, MAID \cite{maid07}, SAID (CM12) \cite{said12}, and J\"uBo \cite{roenchen15} is, in general, quite good.

Our data up to $E_{\gamma} = \unit[930]{MeV}$ were used as a basis for an energy-independent PWA \cite{hartmann14}, allowing for the determination of the $N(1520)\,3/2^-$ helicity amplitudes with minimal model dependence. 
All the data were included in the BnGa multi-channel PWA, together with our recently published data on $G$ \cite{thiel12} and $E$ \cite{gottschall14}, and further data on other channels.%
\footnote{For a complete list, see \cite{hartmann15}, ref. [25] therein.}
Starting from the previous solutions BnGa2011-01 and BnGa2011-02 \cite{anisovich12} all parameters were re-optimized. The newly determined multipoles are compatible with the previous ones at the $2\sigma$ level over the full mass range. The errors are significantly reduced, on average by a factor of 2.25 \cite{hartmann15}. The impact of the new data on the SAID and J\"uBo analyses is currently being investigated in a joint effort of the analysis groups \cite{anisovich16,doering16}.

\subsection{Reaction $\gamma p \to \eta\,p$}
\begin{figure}[p]
	\centering
	\begin{minipage}{\textwidth}
		\includegraphics[width=\textwidth]{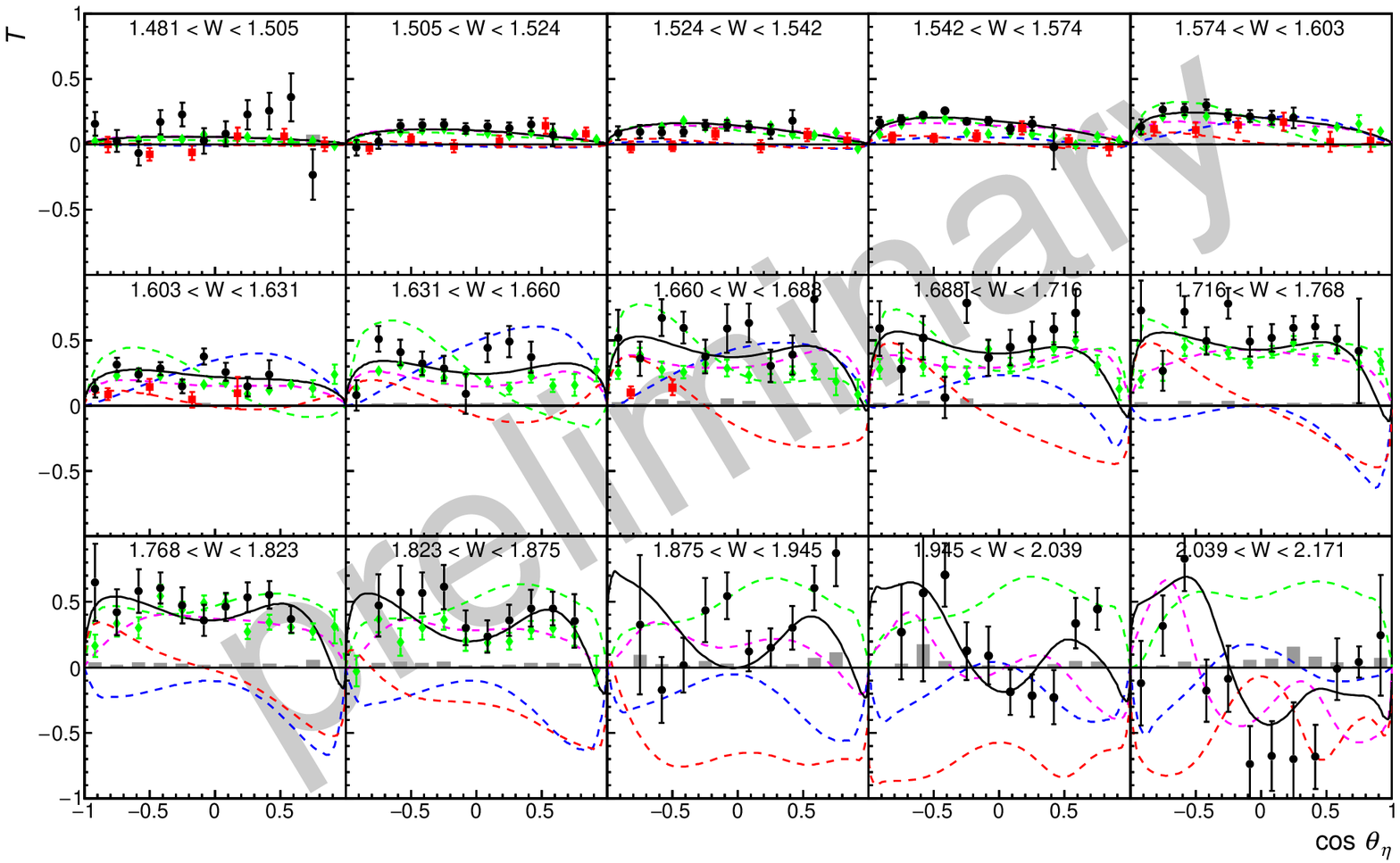}
		\includegraphics[width=\textwidth]{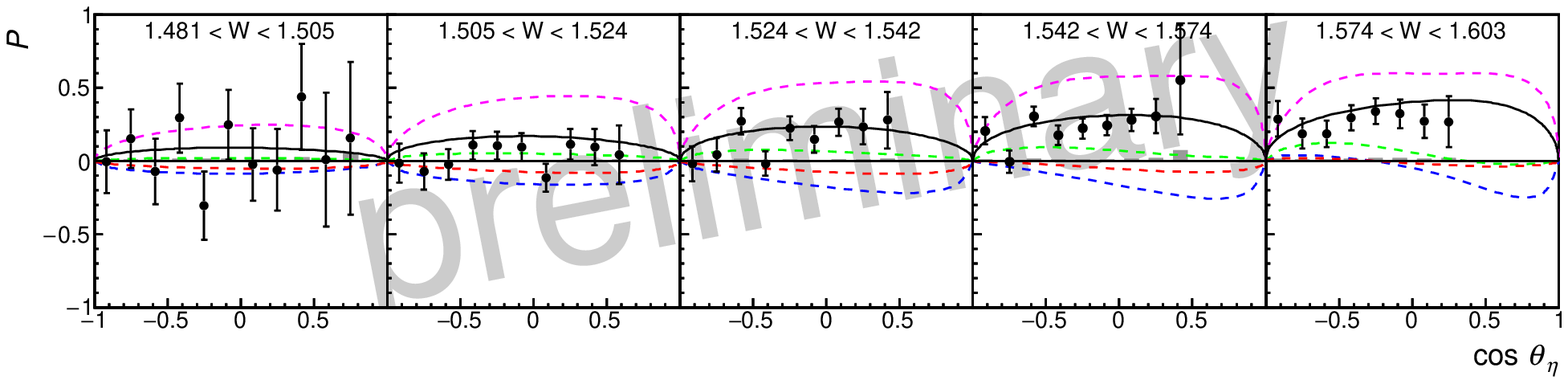}
		\includegraphics[width=\textwidth]{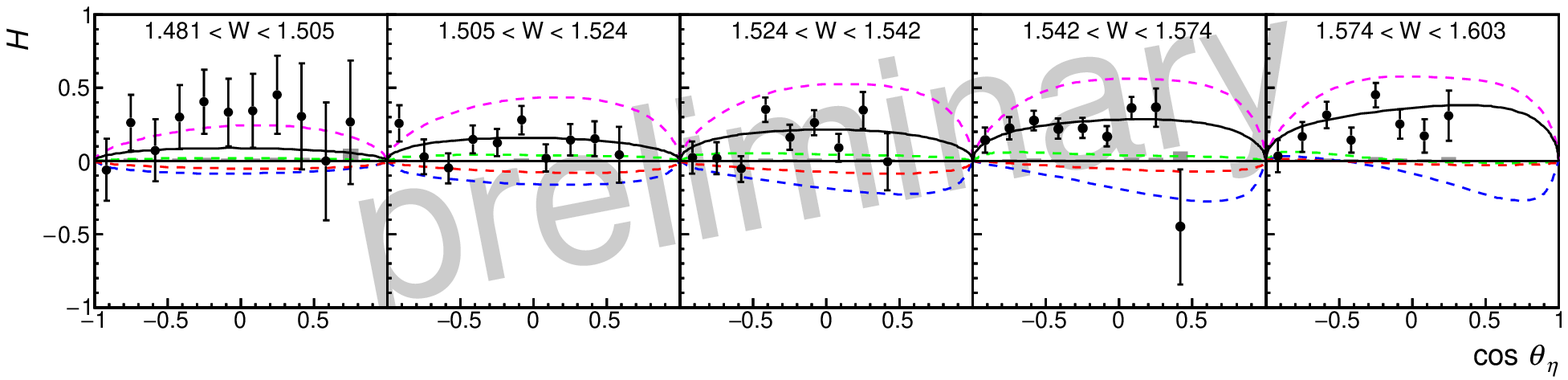}
	\end{minipage}
	\label{fig:results_eta}
	\caption{The polarization observables $T$, $P$, and $H$ in the reaction $\gamma p \to \eta p$ as a function of the scattering angle $\cos\theta_\eta$ and the $\gamma p$ invariant mass $W$ (in \GeV). The systematic uncertainty is shown as gray bars. Earlier ELSA data \cite{bock98} (red) and recent MAMI results \cite{akondi14} (green) are shown for comparison. The solid black line represents a preliminary BnGa fit. The data are compared to predictions (dashed curves) from BnGa2011-02 \cite{anisovich12} (red), MAID \cite{maid07} (green), SAID GE09 \cite{mcnicoll10} (blue), and J\"uBo 2015 \cite{roenchen15} (magenta).}
\end{figure}

Preliminary results for the polarization observables $T$, $P$, and $H$ are shown in Figure~\ref{fig:results_eta}.
Large deviations from the data are observed for the predictions from MAID \cite{maid07}, SAID \cite{mcnicoll10}, BnGa2011 \cite{anisovich12}, Gie\ss{}en \cite{shklyar13}, and the J\"uBo model \cite{roenchen15}, emphasizing how important these new data are to constrain the amplitudes for $\eta$ photoproduction.

The analysis of our new data on $T$, $P$, and $H$, together with not yet published data on $E$ and $G$ and further data from Mainz ($T$, $F$) \cite{akondi14} and JLab \cite{senderovich15} ($E$) by the BnGa group is presently ongoing. The results will be published in the near future \cite{mueller16}.

\section{SUMMARY AND OUTLOOK}
Data have been taken with the CBELSA/TAPS experiment using linearly or circularly polarized photons and a logitudinally or transversely polarized target. In $\pi^0$ photoproduction, the unprecedented precision of the data significantly reduces the errors of the PWA. In $\eta$ photoproduction, where several observables are now measured for the first time, the new data are crucial to constrain the photoproduction amplitudes. Further reaction channels are also being investigated. In particular multi-meson final states like $p \pi^0 \pi^0$ or $p \pi^0 \eta$ are sensitive to cascade decays of higher-mass resonances via intermediate $N^*$ and $\Delta^*$ states \cite{sokhoyan15a,sokhoyan15b,seifen14}. 

We acknowledge support from the \textit{Deutsche Forschungsgemeinschaft} (SFB/TR16) and \textit{Schweizerischer Nationalfonds}.

\bibliographystyle{aipnum-cp}%
\bibliography{hartmann}%

\begin{thebibliography}{31}%
\makeatletter
\providecommand \@ifxundefined [1]{%
 \@ifx{#1\undefined}
}%
\providecommand \@ifnum [1]{%
 \ifnum #1\expandafter \@firstoftwo
 \else \expandafter \@secondoftwo
 \fi
}%
\providecommand \@ifx [1]{%
 \ifx #1\expandafter \@firstoftwo
 \else \expandafter \@secondoftwo
 \fi
}%
\providecommand \natexlab [1]{#1}%
\providecommand \enquote  [1]{``#1''}%
\providecommand \bibnamefont  [1]{#1}%
\providecommand \bibfnamefont [1]{#1}%
\providecommand \citenamefont [1]{#1}%
\providecommand \href@noop [0]{\@secondoftwo}%
\providecommand \href [0]{\begingroup \@sanitize@url \@href}%
\providecommand \@href[1]{\@@startlink{#1}\@@href}%
\providecommand \@@href[1]{\endgroup#1\@@endlink}%
\providecommand \@sanitize@url [0]{\catcode `\$12\catcode `\&12\catcode
  `\#12\catcode `\^12\catcode `\_12\catcode `\%12\relax}%
\providecommand \@@startlink[1]{}%
\providecommand \@@endlink[0]{}%
\providecommand \url  [0]{\begingroup\@sanitize@url \@url }%
\providecommand \@url [1]{\endgroup\@href {#1}{\urlprefix }}%
\providecommand \urlprefix  [0]{URL }%
\providecommand \Eprint [0]{\href }%
\providecommand \doibase [0]{http://dx.doi.org/}%
\providecommand \selectlanguage [0]{\@gobble}%
\providecommand \bibinfo  [0]{\@secondoftwo}%
\providecommand \bibfield  [0]{\@secondoftwo}%
\providecommand \translation [1]{[#1]}%
\providecommand \BibitemOpen [0]{}%
\providecommand \bibitemStop [0]{}%
\providecommand \bibitemNoStop [0]{.\EOS\space}%
\providecommand \EOS [0]{\spacefactor3000\relax}%
\providecommand \BibitemShut  [1]{\csname bibitem#1\endcsname}%
\let\auto@bib@innerbib\@empty
\bibitem [{\citenamefont {Isgur}\ and\ \citenamefont {Karl}(1977)}]{isgur77}%
  \BibitemOpen
  \bibfield  {author} {\bibinfo {author} {\bibfnamefont {N.}~\bibnamefont
  {Isgur}}\ and\ \bibinfo {author} {\bibfnamefont {G.}~\bibnamefont {Karl}},\
  }\href {\doibase 10.1016/0370-2693(77)90074-0} {\bibfield  {journal}
  {\bibinfo  {journal} {Phys. Lett.}\ }\textbf {\bibinfo {volume} {B72}},\ p.\
  \bibinfo {pages} {109} (\bibinfo {year} {1977})}\BibitemShut {NoStop}%
\bibitem [{\citenamefont {Loring}, \citenamefont {Metsch},\ and\ \citenamefont
  {Petry}(2001)}]{loring01}%
  \BibitemOpen
  \bibfield  {author} {\bibinfo {author} {\bibfnamefont {U.}~\bibnamefont
  {Loring}}, \bibinfo {author} {\bibfnamefont {B.~C.}\ \bibnamefont {Metsch}},
  \ and\ \bibinfo {author} {\bibfnamefont {H.~R.}\ \bibnamefont {Petry}},\
  }\href {\doibase 10.1007/s100500170105} {\bibfield  {journal} {\bibinfo
  {journal} {Eur. Phys. J.}\ }\textbf {\bibinfo {volume} {A10}},\ \unskip\
  \bibinfo {pages} {395--446} (\bibinfo {year} {2001})}\BibitemShut {NoStop}%
\bibitem [{\citenamefont {Edwards}\ \emph {et~al.}(2011)\citenamefont
  {Edwards}, \citenamefont {Dudek}, \citenamefont {Richards},\ and\
  \citenamefont {Wallace}}]{edwards11}%
  \BibitemOpen
  \bibfield  {author} {\bibinfo {author} {\bibfnamefont {R.~G.}\ \bibnamefont
  {Edwards}}, \bibinfo {author} {\bibfnamefont {J.~J.}\ \bibnamefont {Dudek}},
  \bibinfo {author} {\bibfnamefont {D.~G.}\ \bibnamefont {Richards}}, \ and\
  \bibinfo {author} {\bibfnamefont {S.~J.}\ \bibnamefont {Wallace}},\ }\href
  {\doibase 10.1103/PhysRevD.84.074508} {\bibfield  {journal} {\bibinfo
  {journal} {Phys. Rev.}\ }\textbf {\bibinfo {volume} {D84}},\ p.\ \bibinfo
  {pages} {074508} (\bibinfo {year} {2011})}\BibitemShut {NoStop}%
\bibitem [{\citenamefont {Chiang}\ and\ \citenamefont
  {Tabakin}(1997)}]{chiang97}%
  \BibitemOpen
  \bibfield  {author} {\bibinfo {author} {\bibfnamefont {W.-T.}\ \bibnamefont
  {Chiang}}\ and\ \bibinfo {author} {\bibfnamefont {F.}~\bibnamefont
  {Tabakin}},\ }\href {\doibase 10.1103/PhysRevC.55.2054} {\bibfield  {journal}
  {\bibinfo  {journal} {Phys. Rev.}\ }\textbf {\bibinfo {volume} {C55}},\
  \unskip\ \bibinfo {pages} {2054--2066} (\bibinfo {year} {1997})}\BibitemShut
  {NoStop}%
\bibitem [{\citenamefont {Thiel}\ \emph {et~al.}(2012)\citenamefont {Thiel}
  \emph {et~al.}}]{thiel12}%
  \BibitemOpen
  \bibfield  {author} {\bibinfo {author} {\bibfnamefont {A.}~\bibnamefont
  {Thiel}} \emph {et~al.},\ }\href {\doibase 10.1103/PhysRevLett.109.102001}
  {\bibfield  {journal} {\bibinfo  {journal} {Phys. Rev. Lett.}\ }\textbf
  {\bibinfo {volume} {109}},\ p.\ \bibinfo {pages} {102001} (\bibinfo {year}
  {2012})}\BibitemShut {NoStop}%
\bibitem [{\citenamefont {Gottschall}\ \emph {et~al.}(2014)\citenamefont
  {Gottschall} \emph {et~al.}}]{gottschall14}%
  \BibitemOpen
  \bibfield  {author} {\bibinfo {author} {\bibfnamefont {M.}~\bibnamefont
  {Gottschall}} \emph {et~al.},\ }\href {\doibase
  10.1103/PhysRevLett.112.012003} {\bibfield  {journal} {\bibinfo  {journal}
  {Phys. Rev. Lett.}\ }\textbf {\bibinfo {volume} {112}},\ p.\ \bibinfo {pages}
  {012003} (\bibinfo {year} {2014})}\BibitemShut {NoStop}%
\bibitem [{\citenamefont {Hillert}(2006)}]{hillert06}%
  \BibitemOpen
  \bibfield  {author} {\bibinfo {author} {\bibfnamefont {W.}~\bibnamefont
  {Hillert}},\ }\href {\doibase 10.1140/epja/i2006-09-015-4} {\bibfield
  {journal} {\bibinfo  {journal} {Eur. Phys. J.}\ }\textbf {\bibinfo {volume}
  {A28S1}},\ \unskip\ \bibinfo {pages} {139--148} (\bibinfo {year}
  {2006})}\BibitemShut {NoStop}%
\bibitem [{\citenamefont {Elsner}\ \emph {et~al.}(2009)\citenamefont {Elsner}
  \emph {et~al.}}]{elsner09}%
  \BibitemOpen
  \bibfield  {author} {\bibinfo {author} {\bibfnamefont {D.}~\bibnamefont
  {Elsner}} \emph {et~al.},\ }\href {\doibase 10.1140/epja/i2008-10708-1}
  {\bibfield  {journal} {\bibinfo  {journal} {Eur. Phys. J.}\ }\textbf
  {\bibinfo {volume} {A39}},\ \unskip\ \bibinfo {pages} {373--381} (\bibinfo
  {year} {2009})}\BibitemShut {NoStop}%
\bibitem [{\citenamefont {Bradtke}, \citenamefont {Dutz}\ \emph
  {et~al.}(1999)\citenamefont {Bradtke}, \citenamefont {Dutz} \emph
  {et~al.}}]{bradtke99}%
  \BibitemOpen
  \bibfield  {author} {\bibinfo {author} {\bibfnamefont {C.}~\bibnamefont
  {Bradtke}}, \bibinfo {author} {\bibfnamefont {H.}~\bibnamefont {Dutz}},
  \emph {et~al.},\ }\href {\doibase 10.1016/S0168-9002(99)00383-6} {\bibfield
  {journal} {\bibinfo  {journal} {Nucl. Instrum. Meth.}\ }\textbf {\bibinfo
  {volume} {A436}},\ \unskip\ \bibinfo {pages} {430--442} (\bibinfo {year}
  {1999})}\BibitemShut {NoStop}%
\bibitem [{\citenamefont {Aker}\ \emph {et~al.}(1992)\citenamefont {Aker} \emph
  {et~al.}}]{aker92}%
  \BibitemOpen
  \bibfield  {author} {\bibinfo {author} {\bibfnamefont {E.}~\bibnamefont
  {Aker}} \emph {et~al.},\ }\href {\doibase 10.1016/0168-9002(92)90379-I}
  {\bibfield  {journal} {\bibinfo  {journal} {Nucl. Instrum. Meth.}\ }\textbf
  {\bibinfo {volume} {A321}},\ \unskip\ \bibinfo {pages} {69--108} (\bibinfo
  {year} {1992})}\BibitemShut {NoStop}%
\bibitem [{\citenamefont {Novotny}(1991)}]{novotny91}%
  \BibitemOpen
  \bibfield  {author} {\bibinfo {author} {\bibfnamefont {R.}~\bibnamefont
  {Novotny}},\ }\href {\doibase 10.1109/23.289329} {\bibfield  {journal}
  {\bibinfo  {journal} {IEEE Trans. Nucl. Sci.}\ }\textbf {\bibinfo {volume}
  {38}},\ \unskip\ \bibinfo {pages} {379--385} (\bibinfo {year}
  {1991})}\BibitemShut {NoStop}%
\bibitem [{\citenamefont {Suft}\ \emph {et~al.}(2005)\citenamefont {Suft} \emph
  {et~al.}}]{suft05}%
  \BibitemOpen
  \bibfield  {author} {\bibinfo {author} {\bibfnamefont {G.}~\bibnamefont
  {Suft}} \emph {et~al.},\ }\href {\doibase 10.1016/j.nima.2004.09.029}
  {\bibfield  {journal} {\bibinfo  {journal} {Nucl. Instrum. Meth.}\ }\textbf
  {\bibinfo {volume} {A538}},\ \unskip\ \bibinfo {pages} {416--424} (\bibinfo
  {year} {2005})}\BibitemShut {NoStop}%
\bibitem [{\citenamefont {Hartmann}\ \emph {et~al.}(2015)\citenamefont
  {Hartmann} \emph {et~al.}}]{hartmann15}%
  \BibitemOpen
  \bibfield  {author} {\bibinfo {author} {\bibfnamefont {J.}~\bibnamefont
  {Hartmann}} \emph {et~al.},\ }\href {\doibase 10.1016/j.physletb.2015.07.008}
  {\bibfield  {journal} {\bibinfo  {journal} {Phys. Lett.}\ }\textbf {\bibinfo
  {volume} {B748}},\ \unskip\ \bibinfo {pages} {212--220} (\bibinfo {year}
  {2015})}\BibitemShut {NoStop}%
\bibitem [{\citenamefont {Hartmann}\ \emph {et~al.}(2016)\citenamefont
  {Hartmann} \emph {et~al.}}]{hartmann16}%
  \BibitemOpen
  \bibfield  {author} {\bibinfo {author} {\bibfnamefont {J.}~\bibnamefont
  {Hartmann}} \emph {et~al.},\ }\href@noop {} {}\ \bibinfo {howpublished}
  {publication in preparation} (\bibinfo {year} {2016} \unskip)\BibitemShut
  {NoStop}%
\bibitem [{\citenamefont {Hartmann}(2016)}]{diss:hartmann}%
  \BibitemOpen
  \bibfield  {author} {\bibinfo {author} {\bibfnamefont {J.}~\bibnamefont
  {Hartmann}},\ }\href@noop {} {\bibinfo {type} {{doctoral thesis in
  preparation}}},\ \bibinfo  {school} {University of Bonn} \bibinfo {year}
  {2016}\BibitemShut {NoStop}%
\bibitem [{\citenamefont {Anisovich}\ \emph {et~al.}(2012)\citenamefont
  {Anisovich} \emph {et~al.}}]{anisovich12}%
  \BibitemOpen
  \bibfield  {author} {\bibinfo {author} {\bibfnamefont {A.~V.}\ \bibnamefont
  {Anisovich}} \emph {et~al.},\ }\href {\doibase 10.1140/epja/i2012-12015-8}
  {\bibfield  {journal} {\bibinfo  {journal} {Eur. Phys. J.}\ }\textbf
  {\bibinfo {volume} {A48}},\ p.~\bibinfo {pages} {15} (\bibinfo {year}
  {2012})}\BibitemShut {NoStop}%
\bibitem [{\citenamefont {Drechsel}, \citenamefont {Kamalov},\ and\
  \citenamefont {Tiator}(2007)}]{maid07}%
  \BibitemOpen
  \bibfield  {author} {\bibinfo {author} {\bibfnamefont {D.}~\bibnamefont
  {Drechsel}}, \bibinfo {author} {\bibfnamefont {S.~S.}\ \bibnamefont
  {Kamalov}}, \ and\ \bibinfo {author} {\bibfnamefont {L.}~\bibnamefont
  {Tiator}},\ }\href {\doibase 10.1140/epja/i2007-10490-6} {\bibfield
  {journal} {\bibinfo  {journal} {Eur. Phys. J.}\ }\textbf {\bibinfo {volume}
  {A34}},\ \unskip\ \bibinfo {pages} {69--97} (\bibinfo {year}
  {2007})}\BibitemShut {NoStop}%
\bibitem [{\citenamefont {Workman}\ \emph {et~al.}(2012)\citenamefont
  {Workman}, \citenamefont {Paris}, \citenamefont {Briscoe},\ and\
  \citenamefont {Strakovsky}}]{said12}%
  \BibitemOpen
  \bibfield  {author} {\bibinfo {author} {\bibfnamefont {R.~L.}\ \bibnamefont
  {Workman}}, \bibinfo {author} {\bibfnamefont {M.~W.}\ \bibnamefont {Paris}},
  \bibinfo {author} {\bibfnamefont {W.~J.}\ \bibnamefont {Briscoe}}, \ and\
  \bibinfo {author} {\bibfnamefont {I.~I.}\ \bibnamefont {Strakovsky}},\ }\href
  {\doibase 10.1103/PhysRevC.86.015202} {\bibfield  {journal} {\bibinfo
  {journal} {Phys. Rev.}\ }\textbf {\bibinfo {volume} {C86}},\ p.\ \bibinfo
  {pages} {015202} (\bibinfo {year} {2012})}\BibitemShut {NoStop}%
\bibitem [{\citenamefont {R{\"o}nchen}\ \emph {et~al.}(2015)\citenamefont
  {R{\"o}nchen} \emph {et~al.}}]{roenchen15}%
  \BibitemOpen
  \bibfield  {author} {\bibinfo {author} {\bibfnamefont {D.}~\bibnamefont
  {R{\"o}nchen}} \emph {et~al.},\ }\href {\doibase 10.1140/epja/i2015-15070-7}
  {\bibfield  {journal} {\bibinfo  {journal} {Eur. Phys. J.}\ }\textbf
  {\bibinfo {volume} {A51}},\ p.~\bibinfo {pages} {70} (\bibinfo {year}
  {2015})}\BibitemShut {NoStop}%
\bibitem [{\citenamefont {Hartmann}\ \emph {et~al.}(2014)\citenamefont
  {Hartmann} \emph {et~al.}}]{hartmann14}%
  \BibitemOpen
  \bibfield  {author} {\bibinfo {author} {\bibfnamefont {J.}~\bibnamefont
  {Hartmann}} \emph {et~al.},\ }\href {\doibase 10.1103/PhysRevLett.113.062001}
  {\bibfield  {journal} {\bibinfo  {journal} {Phys. Rev. Lett.}\ }\textbf
  {\bibinfo {volume} {113}},\ p.\ \bibinfo {pages} {062001} (\bibinfo {year}
  {2014})}\BibitemShut {NoStop}%
\bibitem [{\citenamefont {Anisovich}\ \emph {et~al.}(2016)\citenamefont
  {Anisovich} \emph {et~al.}}]{anisovich16}%
  \BibitemOpen
  \bibfield  {author} {\bibinfo {author} {\bibfnamefont {A.~V.}\ \bibnamefont
  {Anisovich}} \emph {et~al.},\ }\href@noop {} {}\ \bibinfo {howpublished}
  {publication in preparation} (\bibinfo {year} {2016} \unskip)\BibitemShut
  {NoStop}%
\bibitem [{\citenamefont {Doering}(2016)}]{doering16}%
  \BibitemOpen
  \bibfield  {author} {\bibinfo {author} {\bibfnamefont {M.}~\bibnamefont
  {Doering}},\ }\href@noop {} {}\ \bibinfo {howpublished} {Proceedings of the
  XVI International Conference on Hadron Spectorscopy} (\bibinfo {year} {2016}
  \unskip)\BibitemShut {NoStop}%
\bibitem [{\citenamefont {Bock}\ \emph {et~al.}(1998)\citenamefont {Bock} \emph
  {et~al.}}]{bock98}%
  \BibitemOpen
  \bibfield  {author} {\bibinfo {author} {\bibfnamefont {A.}~\bibnamefont
  {Bock}} \emph {et~al.},\ }\href {\doibase 10.1103/PhysRevLett.81.534}
  {\bibfield  {journal} {\bibinfo  {journal} {Phys. Rev. Lett.}\ }\textbf
  {\bibinfo {volume} {81}},\ \unskip\ \bibinfo {pages} {534--537} (\bibinfo
  {year} {1998})}\BibitemShut {NoStop}%
\bibitem [{\citenamefont {Akondi}\ \emph {et~al.}(2014)\citenamefont {Akondi}
  \emph {et~al.}}]{akondi14}%
  \BibitemOpen
  \bibfield  {author} {\bibinfo {author} {\bibfnamefont {C.~S.}\ \bibnamefont
  {Akondi}} \emph {et~al.},\ }\href {\doibase 10.1103/PhysRevLett.113.102001}
  {\bibfield  {journal} {\bibinfo  {journal} {Phys. Rev. Lett.}\ }\textbf
  {\bibinfo {volume} {113}},\ p.\ \bibinfo {pages} {102001} (\bibinfo {year}
  {2014})}\BibitemShut {NoStop}%
\bibitem [{\citenamefont {McNicoll}\ \emph {et~al.}(2010)\citenamefont
  {McNicoll}, \citenamefont {Prakhov}, \citenamefont {Strakovsky} \emph
  {et~al.}}]{mcnicoll10}%
  \BibitemOpen
  \bibfield  {author} {\bibinfo {author} {\bibfnamefont {E.~F.}\ \bibnamefont
  {McNicoll}}, \bibinfo {author} {\bibfnamefont {S.}~\bibnamefont {Prakhov}},
  \bibinfo {author} {\bibfnamefont {I.~I.}\ \bibnamefont {Strakovsky}},  \emph
  {et~al.},\ }\href {\doibase 10.1103/PhysRevC.82.035208} {\bibfield  {journal}
  {\bibinfo  {journal} {Phys. Rev.}\ }\textbf {\bibinfo {volume} {C82}},\ p.\
  \bibinfo {pages} {035208} (\bibinfo {year} {2010})}\BibitemShut {NoStop}%
\bibitem [{\citenamefont {Shklyar}, \citenamefont {Lenske},\ and\ \citenamefont
  {Mosel}(2013)}]{shklyar13}%
  \BibitemOpen
  \bibfield  {author} {\bibinfo {author} {\bibfnamefont {V.}~\bibnamefont
  {Shklyar}}, \bibinfo {author} {\bibfnamefont {H.}~\bibnamefont {Lenske}}, \
  and\ \bibinfo {author} {\bibfnamefont {U.}~\bibnamefont {Mosel}},\ }\href
  {\doibase 10.1103/PhysRevC.87.015201} {\bibfield  {journal} {\bibinfo
  {journal} {Phys. Rev.}\ }\textbf {\bibinfo {volume} {C87}},\ p.\ \bibinfo
  {pages} {015201} (\bibinfo {year} {2013})}\BibitemShut {NoStop}%
\bibitem [{\citenamefont {Senderovich}\ \emph {et~al.}(2015)\citenamefont
  {Senderovich} \emph {et~al.}}]{senderovich15}%
  \BibitemOpen
  \bibfield  {author} {\bibinfo {author} {\bibfnamefont {I.}~\bibnamefont
  {Senderovich}} \emph {et~al.},\ }\href@noop {} {\  (\bibinfo {year}
  {2015})},\ \Eprint {http://arxiv.org/abs/1507.00325} {arXiv:1507.00325
  [nucl-ex]} \BibitemShut {NoStop}%
\bibitem [{\citenamefont {M{\"u}ller}\ \emph {et~al.}(2016)\citenamefont
  {M{\"u}ller}, \citenamefont {Hartmann}, \citenamefont {Gr{\"u}ner} \emph
  {et~al.}}]{mueller16}%
  \BibitemOpen
  \bibfield  {author} {\bibinfo {author} {\bibfnamefont {J.}~\bibnamefont
  {M{\"u}ller}}, \bibinfo {author} {\bibfnamefont {J.}~\bibnamefont
  {Hartmann}}, \bibinfo {author} {\bibfnamefont {M.}~\bibnamefont
  {Gr{\"u}ner}},  \emph {et~al.},\ }\href@noop {} {}\ \bibinfo {howpublished}
  {publication in preparation} (\bibinfo {year} {2016} \unskip)\BibitemShut
  {NoStop}%
\bibitem [{\citenamefont {Sokhoyan}\ \emph
  {et~al.}(2015{\natexlab{a}})\citenamefont {Sokhoyan} \emph
  {et~al.}}]{sokhoyan15a}%
  \BibitemOpen
  \bibfield  {author} {\bibinfo {author} {\bibfnamefont {V.}~\bibnamefont
  {Sokhoyan}} \emph {et~al.},\ }\href {\doibase 10.1016/j.physletb.2015.04.063}
  {\bibfield  {journal} {\bibinfo  {journal} {Phys. Lett.}\ }\textbf {\bibinfo
  {volume} {B746}},\ \unskip\ \bibinfo {pages} {127--131} (\bibinfo {year}
  {2015}{\natexlab{a}})}\BibitemShut {NoStop}%
\bibitem [{\citenamefont {Sokhoyan}\ \emph
  {et~al.}(2015{\natexlab{b}})\citenamefont {Sokhoyan} \emph
  {et~al.}}]{sokhoyan15b}%
  \BibitemOpen
  \bibfield  {author} {\bibinfo {author} {\bibfnamefont {V.}~\bibnamefont
  {Sokhoyan}} \emph {et~al.},\ }\href {\doibase 10.1140/epja/i2015-15095-x}
  {\bibfield  {journal} {\bibinfo  {journal} {Eur. Phys. J.}\ }\textbf
  {\bibinfo {volume} {A51}},\ p.~\bibinfo {pages} {95} (\bibinfo {year}
  {2015}{\natexlab{b}})}\BibitemShut {NoStop}%
\bibitem [{\citenamefont {Seifen}(2014)}]{seifen14}%
  \BibitemOpen
  \bibfield  {author} {\bibinfo {author} {\bibfnamefont {T.}~\bibnamefont
  {Seifen}},\ }\href {\doibase 10.1142/S2010194514600659} {\bibfield  {journal}
  {\bibinfo  {journal} {Int. J. Mod. Phys. Conf. Ser.}\ }\textbf {\bibinfo
  {volume} {26}},\ p.\ \bibinfo {pages} {1460065} (\bibinfo {year}
  {2014})}\BibitemShut {NoStop}%
\end{thebibliography}%

\end{document}